\documentclass [twocolumn,superscriptaddress]{revtex4}

\usepackage {graphicx}
\usepackage {amsmath}
\usepackage {bm}
\usepackage {subfigure}
\usepackage {hyperref}
\newsavebox{\mylegend}
\sbox{\mylegend}
{\includegraphics[width=0.15\textwidth,bb= 130   400   250   540]{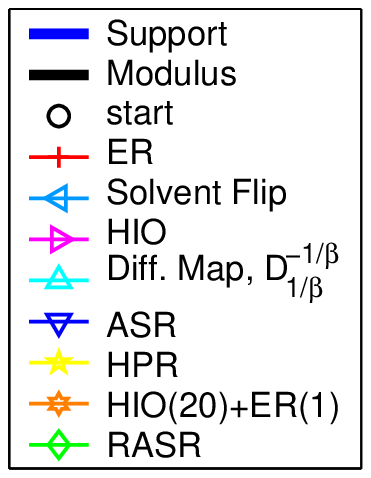}}

\begin{document}

\title{Benchmarking Iterative Projection Algorithms for Phase Retrieval}

\begin{abstract}
Iterative projection algorithms for phase retrieval are tested on two simple
`toy' models. The result provides useful insights in the behavior of these
algorithms.
\end{abstract}

\author{S.~Marchesini}
\noaffiliation
\email[Correspondence and requests for materials should be addressed
to S. Marchesini: ]{smarchesini@llnl.gov}
 \affiliation{Lawrence Livermore National Laboratory, 7000 East Ave.,
Livermore, CA 94550-9234, USA}
%\pacs{61.10.Nz  42.30.Rx  42.30.Wb  68.37.Yz}

%\preprint{UCRL-JC-153571}

\date{\today}
\maketitle
%%%%%%%%%%%%%%%%%%%%%
Iterative transform methods pioneered by Gerchberg and Saxton
\cite{Gerchberg:1972},  are well established techniques for
iteratively recovering the phase from the knowledge of the diffraction
amplitude.  The development of iterative algorithms with feedback in
the early nineteen-eighties by Fienup produced a remarkably successful
optimization method capable of extracting phase information from
adequately sampled intensity data \cite{fienup:1978,fienup:1982, cederquist:1988}. 
Finally, the important theoretical insight that these iterations may 
be viewed as projections in Hilbert space 
\cite{stark:1984,stark:1987} has allowed theoreticians to analyze and 
improve on the basic Fienup algorithm
\cite{elser:2003, luke:1,luke:2,luke:3}.

These algorithms try to find the intersection between two sets, typically
the set of all the possible objects with a given diffraction pattern
(modulus), and the set of all the objects that are constrained within a
given area called support (or solvent in crystallography).  
The search for the intersection is based on
the information obtained by `projecting' the current estimate on the two
sets. An error metric exists to characterize the distance between the
current estimate and a given feasibility set. The error metric and 
its gradient are used in conjugate gradient (CG) based methods 
such as SPEDEN \cite{speden}.
A projector $\bm{P}$ is an operator that takes to the closest 
point of a set from the current point $\rho$. 
A repetition of the same projection is equal to one
projection alone ($\bm{P^2}=\bm{P}$). 
Another operator used here is the reflector
$\bm{R}=2\bm{P}-\bm{I}$. 
 We consider two sets, $S$ (support) and $M$
(modulus). The support constraint is convex, while the modulus constraint
is non-convex. Problems arise for non-convex sets, 
where projections become multivalued \cite{luke:siam}.

The support projector $\bm{P}_s$ acts on the object density $\rho$ by 
setting to 0 the density of the object outside a given region.
The modulus projector $\bm{P}_m$ acts on the density $\rho$ in the 
Fourier domain $\tilde \rho$ by forcing the modulus $|\tilde \rho|$ to be 
equal to the known one $m$, but keeping the phase of the current 
object in the Fourier domain $\tilde \rho$.
This operator is demonstrated to be a
 projector on the non-covex set of the magnitude constraint
 \cite{luke:siam}.
The same paper discusses the problems of multi-valued projections 
for non-convex sets, which do not statisfy the requirements 
for gradient-based minimization algorithms, 
and the related nonsmoothness of the squared set 
distance metric, which may lead to numerical instabilities.  
See also \cite{luke:siam1} for a follow-up discussion on the 
non-smooth analysis. 

Several algorithms based  on these concepts have now been proposed 
and a visual representation of their behaviour is usefull to characterize
the algorithm in various situations, in order to help chose the most 
appropriate one for a particular problem. 

The following algorithms require a starting point $\rho^0$, 
which is generated by 
assigning a random phase to the measured object amplitude in the Fourier 
domain $|\tilde \rho|$.
The first algorithm called {\it Error Reduction} (ER) (Gerchberg and Saxton
\cite{Gerchberg:1972}) (see also Alternating Projections Onto Convex Sets 
\cite{bregman:1965} or Alternating Projections Onto Nononvex Sets 
 \cite{stark:1984}) is simply:
\begin{equation}
\rho^{(n+1)}=\bm{P_s P_m}\rho^{(n)\,,}
\end{equation}
by projecting back and forth between two sets, it converges to the local
minimum (gradient type).  
The eigenvalues of the support projectors are 0 and 1, with corrisponding
eigenvectors the pixels outside and inside the support. Replacing
the support projector  $\bm{P}_s$ with its reflector 
$\bm{R}_s=2\bm{P}_s-\bm{I}$, the
corrisponding eigenvalues become -1 and 1, i.e. the charge 
density $\rho$ outside the support is multiplied by -1.
This algorithm is called {\it solvent flipping} in crystallography 
\cite{abrahams:1996}:
\begin{equation}
\rho^{(n+1)}=\bm{R_s P_m}\rho^{(n)} \,.
\end{equation}

\noindent
The {\it Hybrid Input Output} (HIO) \cite{fienup:1978,fienup:1982} is
\begin{equation}
\label{eq:HIO}
 \rho^{(n+1)}(x)=
\begin{cases}
\bm{P_m} \rho^{(n)}(x)  & 
	\text {if  $x\in S$} \\
(\bm{I}-\beta \bm{P_m})\rho^{(n)}(x)  & \text{otherwise}
\end{cases}
\end{equation}
%\begin{equation}
%\rho^{(n+1)}=[\bm{P_s P_m}+(\bm{I} - \bm{P_s})(1-\beta \bm{P_m})]\rho^{(n)} 
%\end{equation}
\noindent
It is often used in conjunction of the ER algorithm, alternating
several HIO iterations and one ER iteration (HIO(20)+ER(1) in our case).
{\it Difference Map} with $\gamma_1=-\beta^{-1}$, $\gamma_2=\beta^{-1}$
\cite{elser:2003},  which requires 4 projections (two
time-consuming modulus constraint projections):
\begin{eqnarray}
\nonumber
\rho^{(n+1)}=
\{&
	\bm{I}&+%\\ \nonumber
	\bm{P_S} 
 \left [
  \left (
	\beta+1
  \right )
  \bm{P_m}-\bm{I}
 \right ]\\
&-&\bm{P_m} 
 \left [
  \left (
	\beta-1
  \right )
  \bm{P_s}+\bm{I}
 \right ]
\}
\rho^{(n)} 
\end{eqnarray}
The {\it Averaged Successive Reflections} (ASR) \cite{luke:1} is:
\begin{equation}
\rho^{(n+1)}=\tfrac{1}{2}[\bm{R_s R_m}+\bm{I} ]\rho^{(n)} 
\end{equation}
The {\it Hybrid Projection Reflection} (HPR) \cite{luke:2} 
 is derived from a relaxation of ASR:
\begin{eqnarray}
\nonumber
\rho^{(n+1)}&=&[
	\bm{R_s} 
	\left (
		\bm{R_m}+(\beta-1) \bm{P_m} 
	\right )\\
	&+&\bm{I}
	+(1-\beta )\bm{P_m}
	]
\rho^{(n)} 
\end{eqnarray}
It is equivalent to HIO if positivity is not enforced 
but it is written in a recursive form, instead of a 
case by case form such as Eq. \ref{eq:HIO}.
Finally {\it Relaxed Averaged Alternating Reflectors} RAAR (previously
named RARS) \cite{luke:3}
\begin{equation}
\rho^{(n+1)}=\left [ \tfrac{1}{2} \beta \left (
	\bm{R_s R_m}+\bm{I} 
	\right )
	+(1-\beta)\bm{P_m}
	\right ] 
\rho^{(n)} 
\end{equation}
For $\beta=1$, HIO, HPR, ASR and RAAS coincide.

\begin{figure*}[htbp]
\subfigure[]
    {
      \label{2linesa}
	\includegraphics[width=0.4\textwidth]{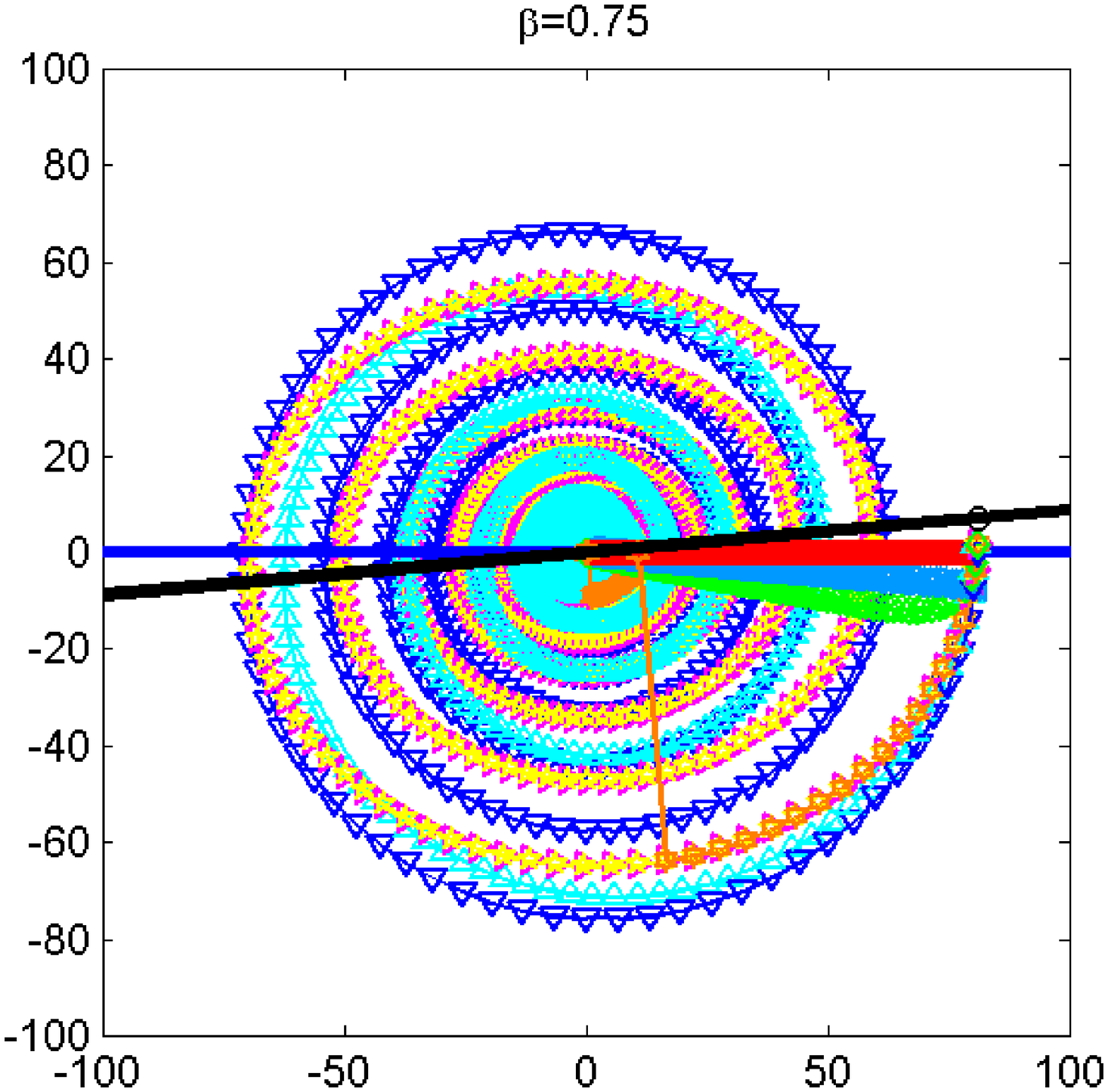}
	\usebox{\mylegend}
    }
\subfigure[]
    {
      \label{2linesb}
		\includegraphics[width=0.4\textwidth]{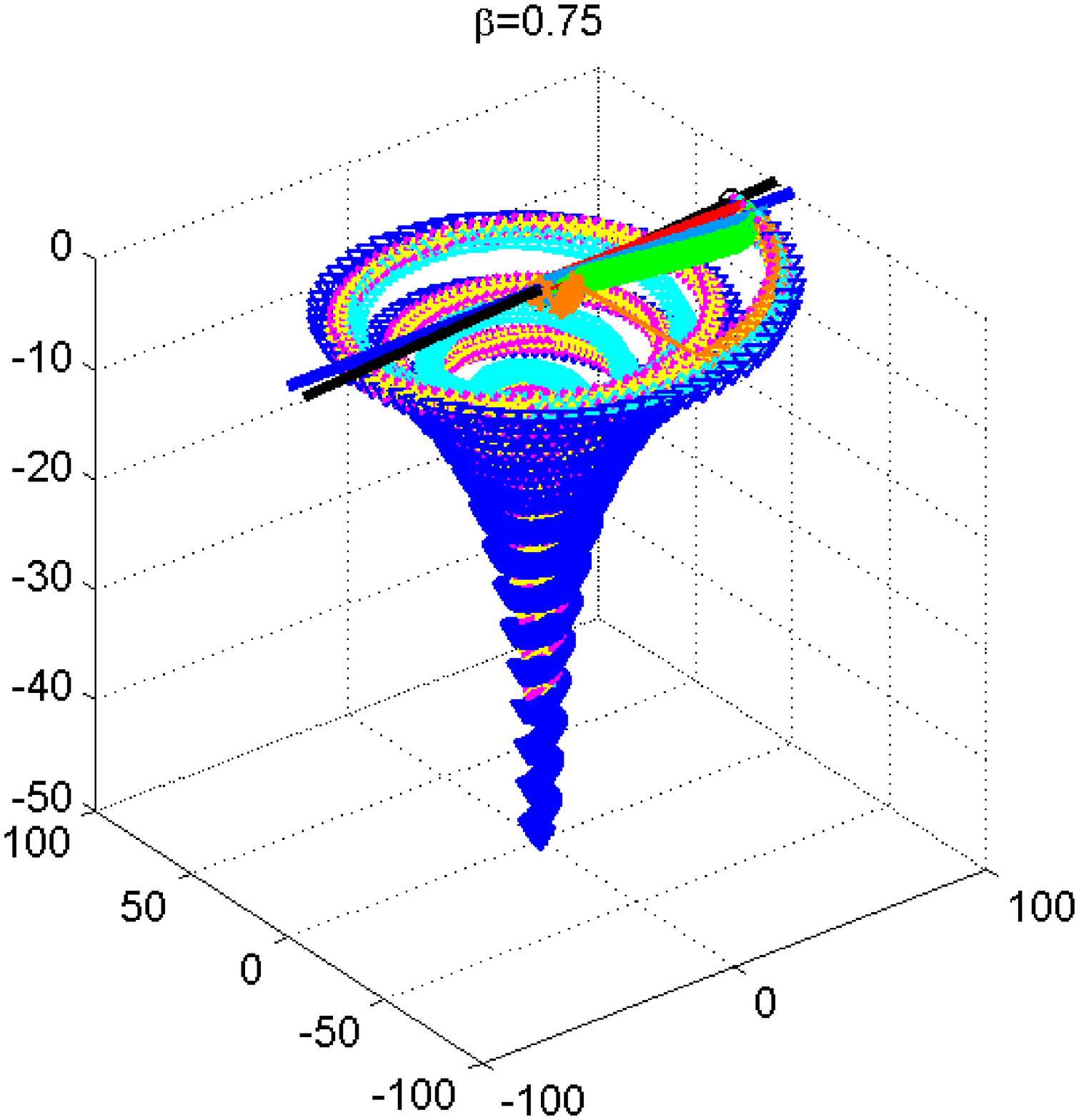}
    }
\caption{
The basic features of the iterative projection algorithms can understood by
this simple model of two lines intersecting (\ref{2linesa}). 
The aim is to find the
intersection.  The ER algorithm and the Solvent flipping algorithms converge in
some gradient type fashion (the distance to the two sets never increases), 
with the solvent flip method being slightly faster
when the angle between the two lines is small. 
HIO and variants move slightly in the direction where the gap 
between the two projections decreases,
but at the same time in the direction of the gap, following a spiral path. 
When the two lines do not intersect (\ref{2linesb}, HIO and
variants keep moving in the direction of the gap. ER, Solvent Flipping and
 RAAR  converge at (or close to) the local minimum.}
\label{2lines}
\end{figure*}

The first test is performed on the simplest possible case: find the
intersection between two lines. Fig. \ref{2lines} shows the behavior
of the various algorithms, The two sets are represented by a horizontal
blue line (support) and a tilted black line (modulus). 
ER simply projects back and
forth between these two lines, and moves along the support line in 
the direction of the intersection. 
Solvent Flip projects onto the modulus,  `reflects' on the support, 
and moves along the reflection of the modulus constraint onto the support. 
The solvent flipping algorithm is slightly faster than ER due
to the increase in the angle of the projections and reflections.
HIO and variants (ASR, Difference Map,
HPR and RAAS) move in a spiral around the intersection eventually
reaching the intersection. For similar $\beta$ RAAS behaves somewhere in
between ER and HIO with a sharper spiral, reaching the solution much
earlier. Alternating 20 iteration of HIO and 1 of ER (HIO(20)+ER(1)) 
considerably speeds up convergence. 

When a gap is introduced between the two lines (Fig. \ref{2linesb})
 so that the two lines do not intersect, HIO and variants move
away from this local minimum in search for another `attractor' or local
minimum. This shows how these algorithms escape from local minima and explore
the Hilbert space for other minima. ER, Solvent Flip, 
RAAS converges to or near the local minimum. 
By varying $\beta$ RAAS becomes a local minimizer for small $\beta$, and
becomes like HIO for $\beta\simeq 1$. ER, solvent flip HIO+ER converge
to the local minimum. The properties of SPEDEN cannot be fully apreciated 
in these examples since the support constraint is represented by a one 
dimensional set, and this conjugate gradient method is designed for 
multidimensional minimization.  However it is important to remark that 
such algorithm converges quadratically to the local minimum, reaching 
the intersection in a single step when it exists, 
and the local minimum when a gap is introduced between the 
two sets in fewer steps than any of the other 
algorithm described here.

\begin{figure*}[htbp]
\subfigure[]
    {\label{2circlesa}
	\includegraphics[width=0.32\textwidth]{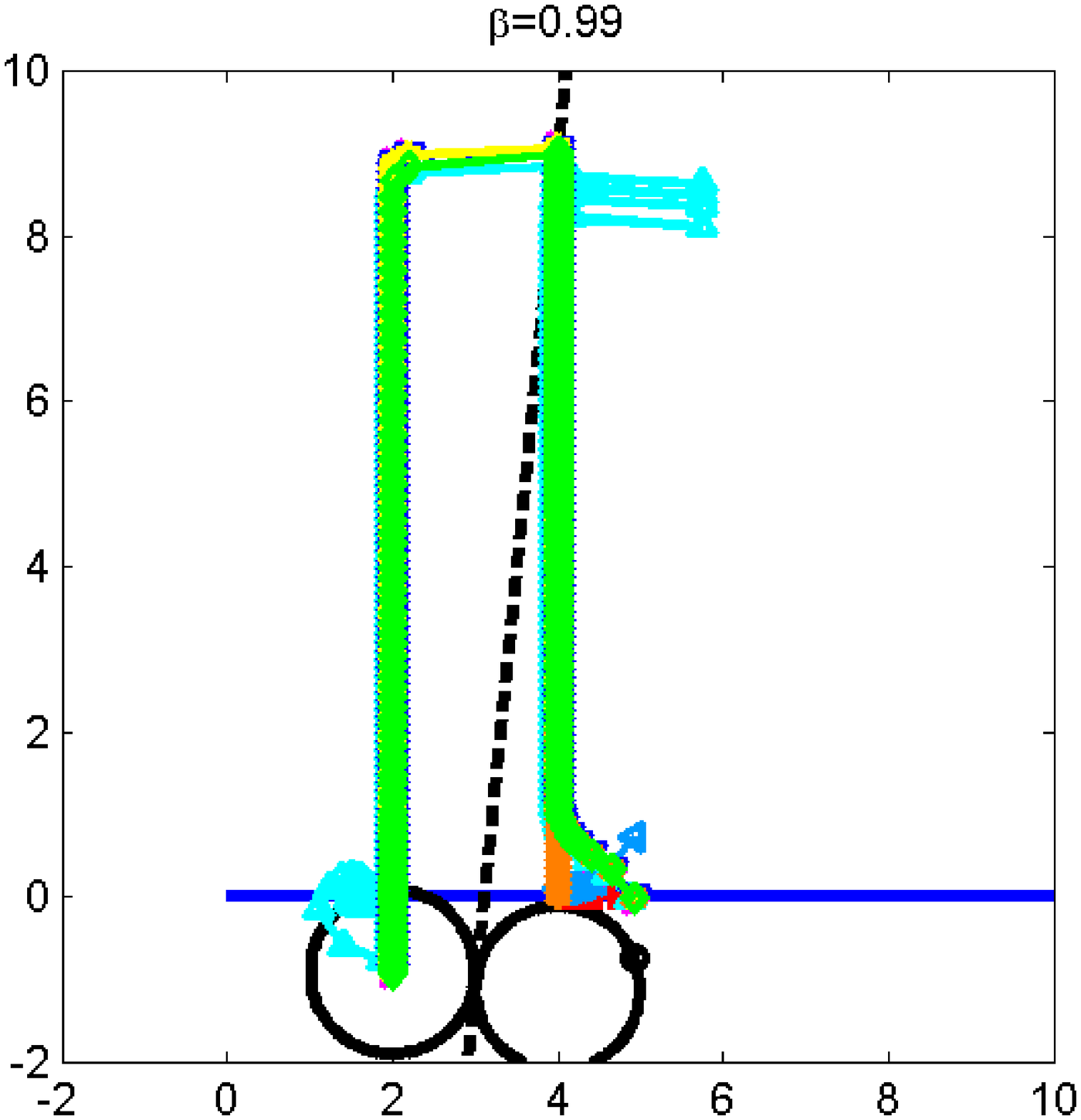}
	\usebox{\mylegend}
	\includegraphics[width=0.32\textwidth]{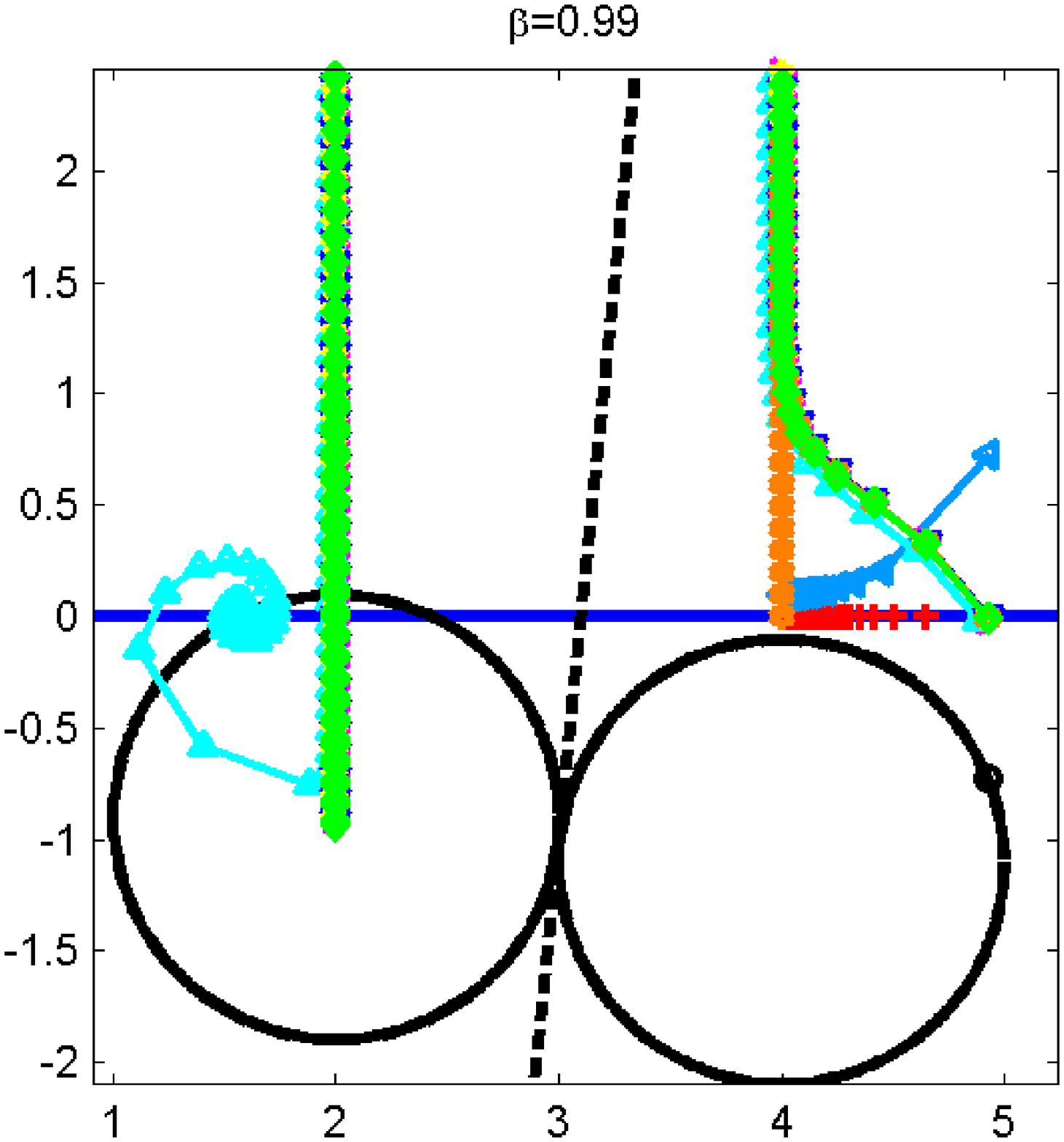}
	}\\[-19pt]
\subfigure[]
    {\label{2circlesb}
\includegraphics[width=0.32\textwidth]{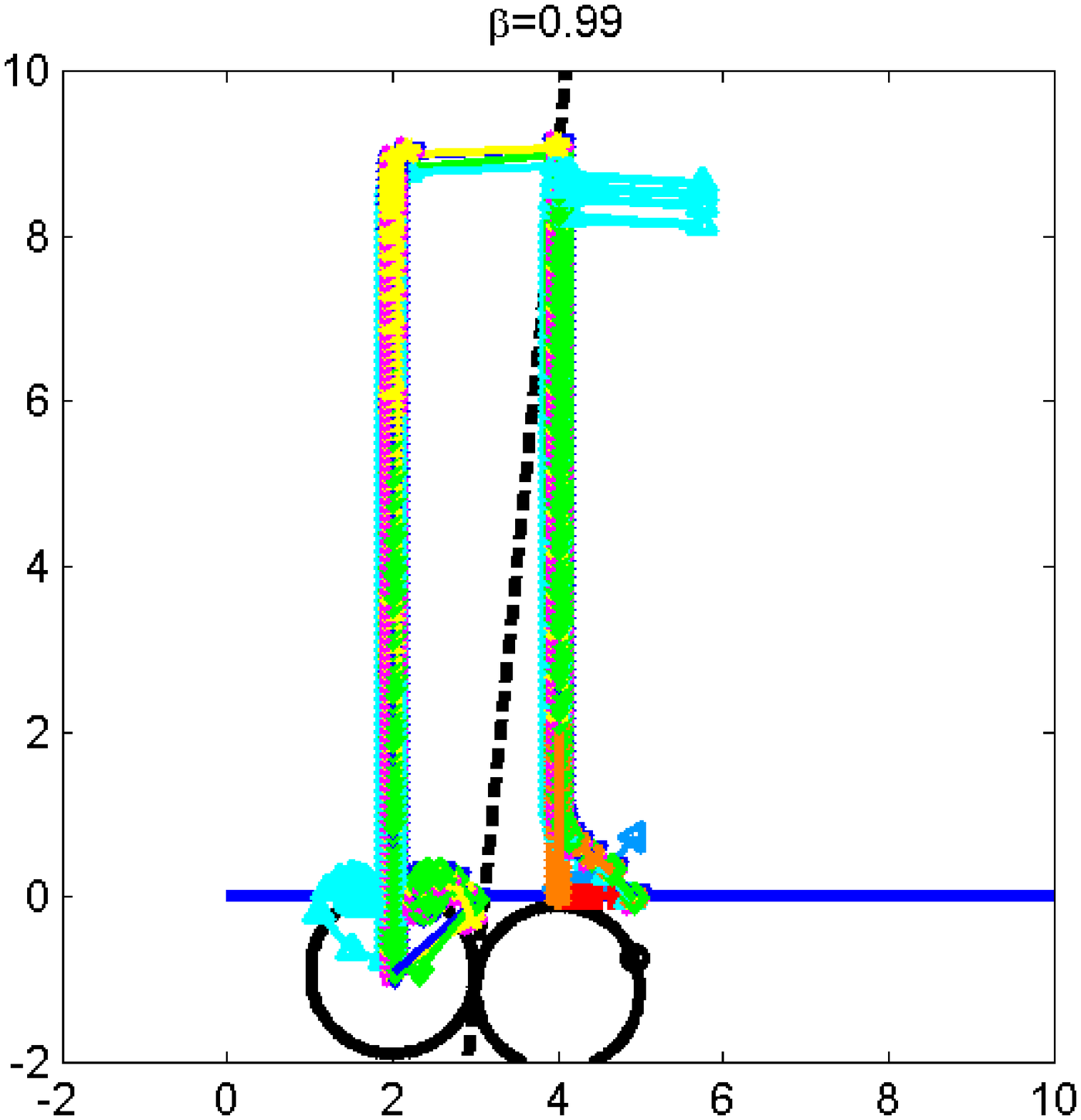}
	\usebox{\mylegend}
\includegraphics[width=0.32\textwidth]{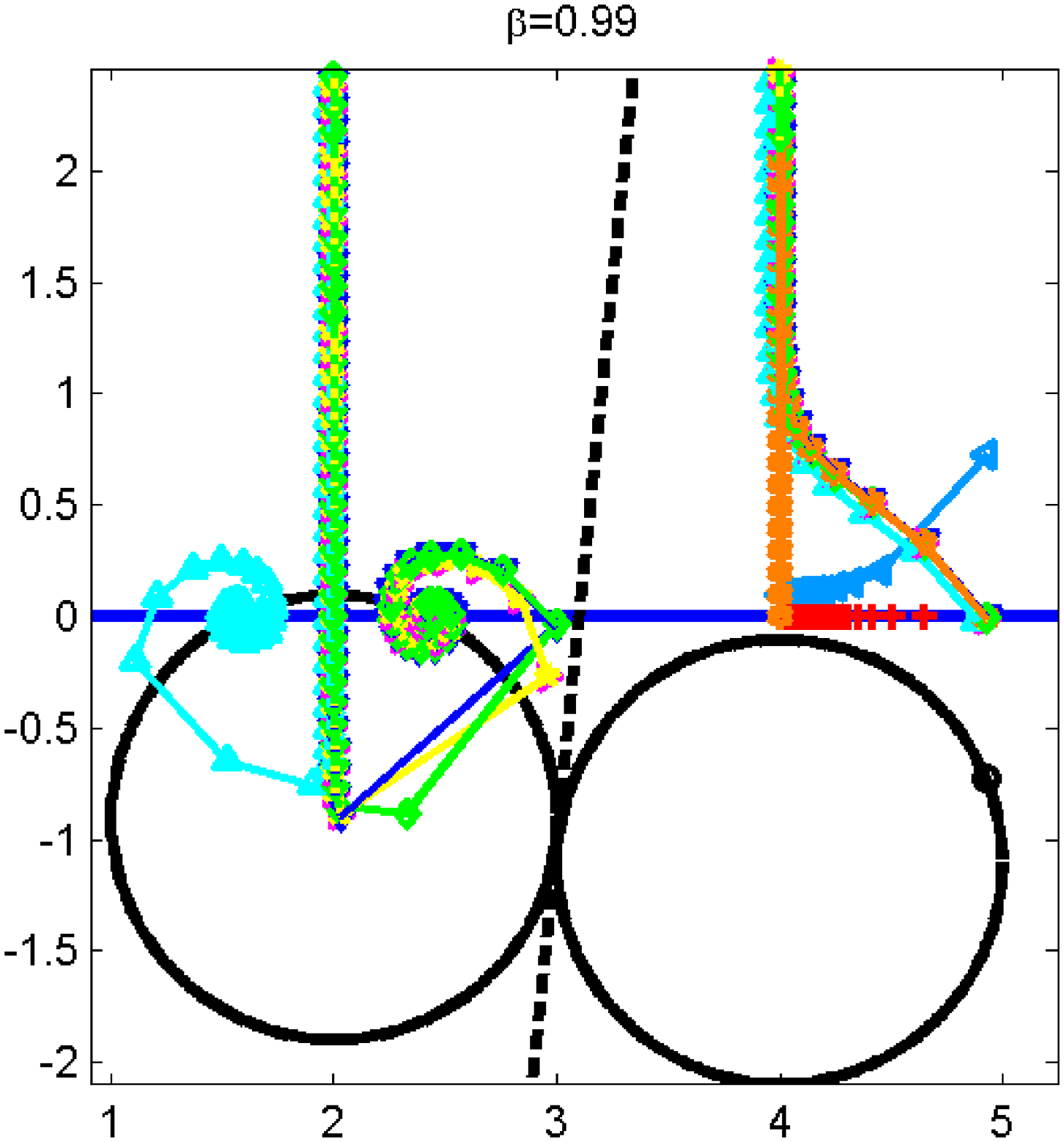}
}\\[-19pt]
%\subfigure[]
%    {\label{2circlesc}
%\includegraphics[width=0.32\textwidth]{2circlessmb.eps}
%	\usebox{\mylegend}
%\includegraphics[width=0.32\textwidth]{2circlessmb_z.eps}
%}\\[-19pt]
\subfigure[]
    {\label{2circlesd}
\includegraphics[width=0.32\textwidth]{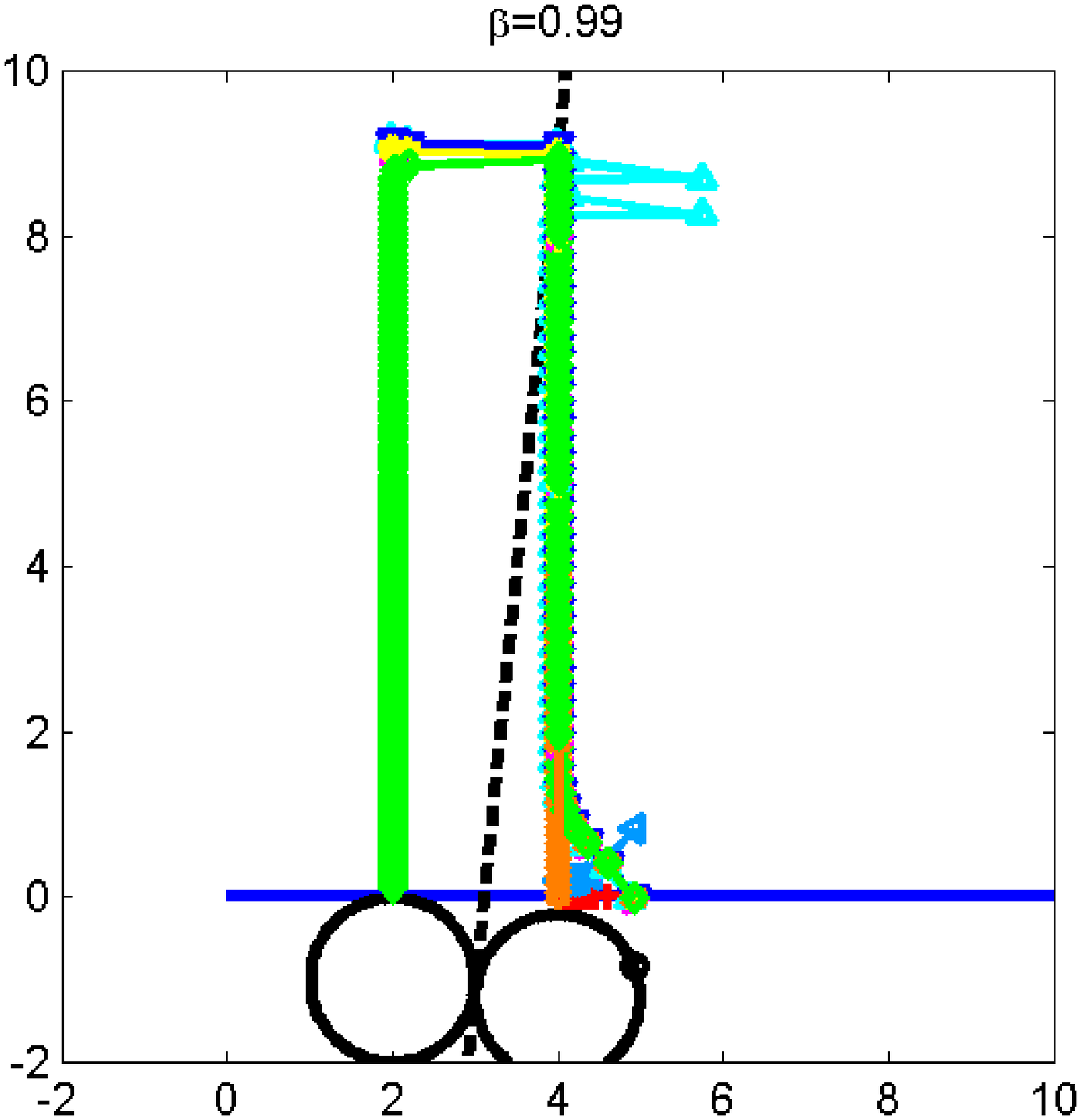}
	\usebox{\mylegend}
\includegraphics[width=0.32\textwidth]{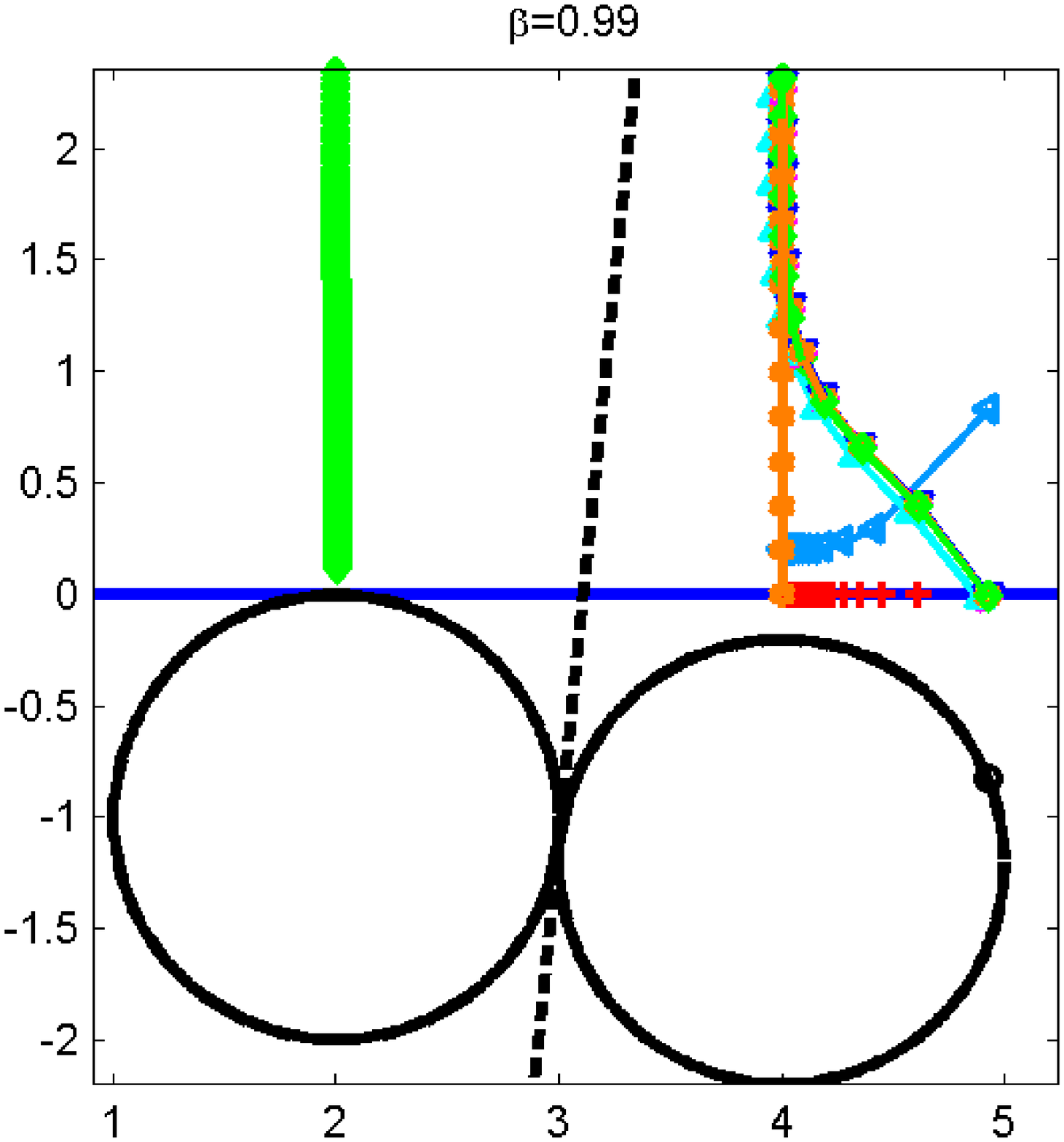}
}\\[-10pt]
\caption{The horizontal line represents a support constraint, 
while the two circles represent a non convex constraint, i.e. 
the modulus constraint. 
The dashed line divides 
the region closer to one circle from the other.
The starting point is on the circle to the right, 
possessing a local minimum distance to the line. 
(a) The gradient-type (ER and Solvent Flip) algorithms converge 
to the local minimum,
while HIO and variants move away from the local minimum in the direction 
of the gap (vertical) untill they reach the region where the second 
circle is closer (delimited by the dashed line).  
From here they try to move in the same spiral-like path of the 
two lines (Fig. \ref{2lines}) untill they reach the point where the 
projecton on the circle and the line are parallel, 
and start moving toward the the center of the circle which has 
the correct solution. They stagnate in the center of the circle where 
the projection is multivalued.
Only the Diff. Map reaches one of the two solutions. 
The addition of a small value  of the order of the numerical precision
 after each iteration  solves this stagnation (b). 
When one of the circles just touches the other constraint most algorithms
either get stuck near the local minimum or stagnate. RAAS is the only
one that reaches the vicinity of the solution (c).
}
\label{2circles}
\end{figure*}

A more realistic example is shown in Fig. \ref{2circles}. 
Here the circumference of two circles represent the modulus 
constraint, while the support
constraint is represented by a line. The two circles are used to 
represent a non-convex set with a local minimum.
It is difficult to represent a true modulus constraint in real 
space. For a representation of the modulus 
constraint in reciprocal space see \cite{luke:siam}.
The advantage of this example
is the simplicity in the `modulus' projector operator (it projects onto 
the closest circle). 
Although a real modulus constraint projector is not as simple as 
the one used in this example,
there are similarities: each Fourier space  point provides an
n-dimensional ellipsoid type equation. 

We start from a position near the local minimum. ER, solvent flip and
HIO+ER all fall into this trap (Fig. \ref{2circlesa}), although
increasing the interval between ER iterations in the HIO+ER algorithm
would allow it to escape this local minimum. 
HIO and variants move away from the local minimum, `find' the other
circle, but converge to the center of the circle, with all but
Diff. Map. not reaching a solution. In the center of the circle the
projection on the modulus constraint becomes `multivalued', 
and its distance metric is `nonsmooth'.
The introduction of a small a random number added to the resulting
solution at every step allows all the HIO-type codes to escape stagnation 
and find the solution (Fig. \ref{2circlesb}). 
The random number can be as low as the numerical precision of the computer. 
For $\beta$ reduced to $.9$, RAAS would not reach the solution, but
converge close to the local  minimum. 
As a latest test in this series Fig. \ref{2circlesd}, 
 shows the behaviour of the algorithms when the support is tangent to
the circle, the two solutions coincide, and the the two constraints 
are parallel.
The only algorithm to reach the solution is RAAS, but HIO+ER would
also reach the solution if the interval between ER steps was sufficiently
large. 

%%%%%%%%%%%%%%%%%%%55

\section{Positivity}
\begin{figure*}[tbp]
\subfigure[]
{\label{2circlesposa}
\includegraphics[width=0.35\textwidth]{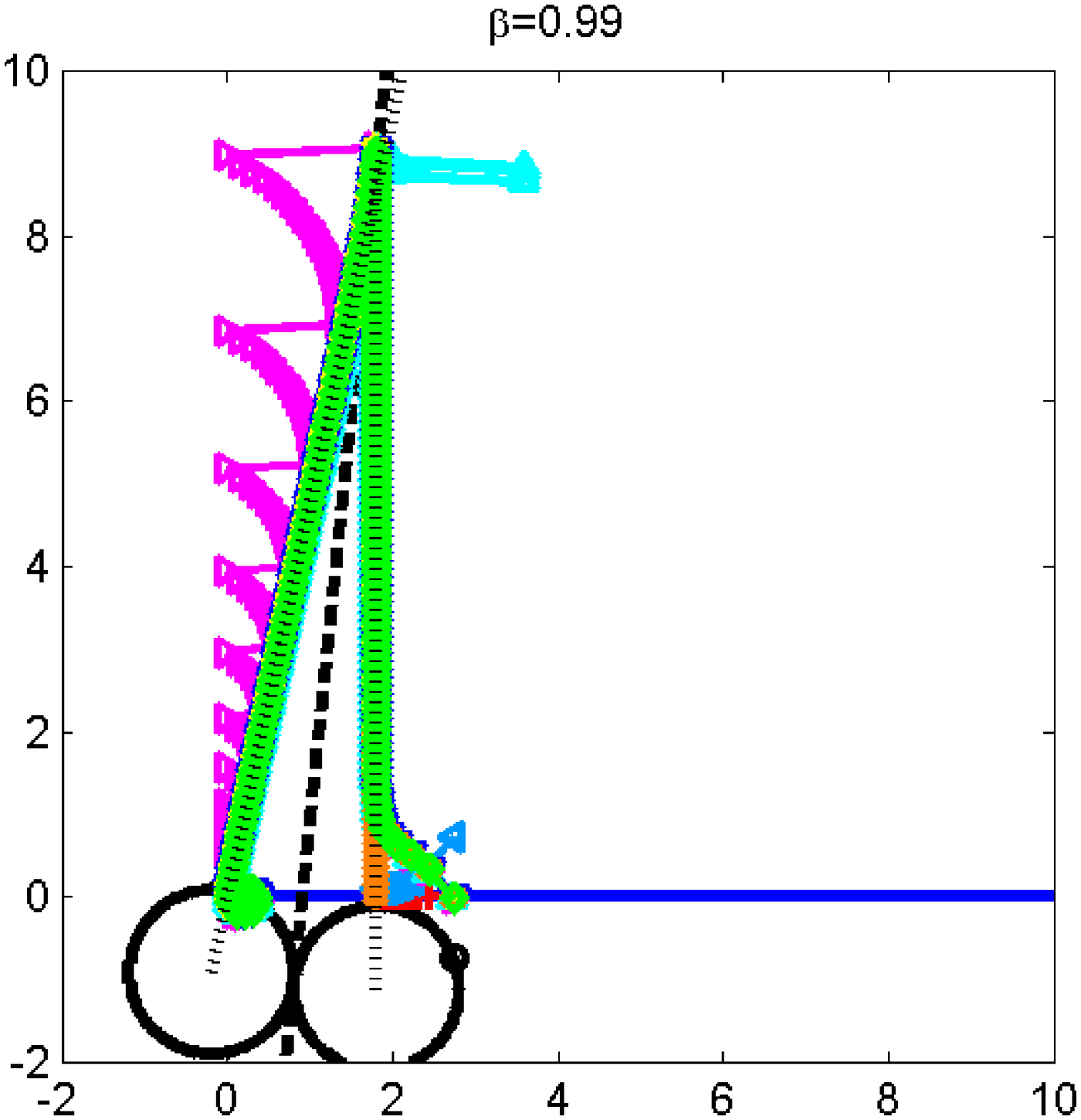}
\usebox{\mylegend}
\includegraphics[width=0.35\textwidth]{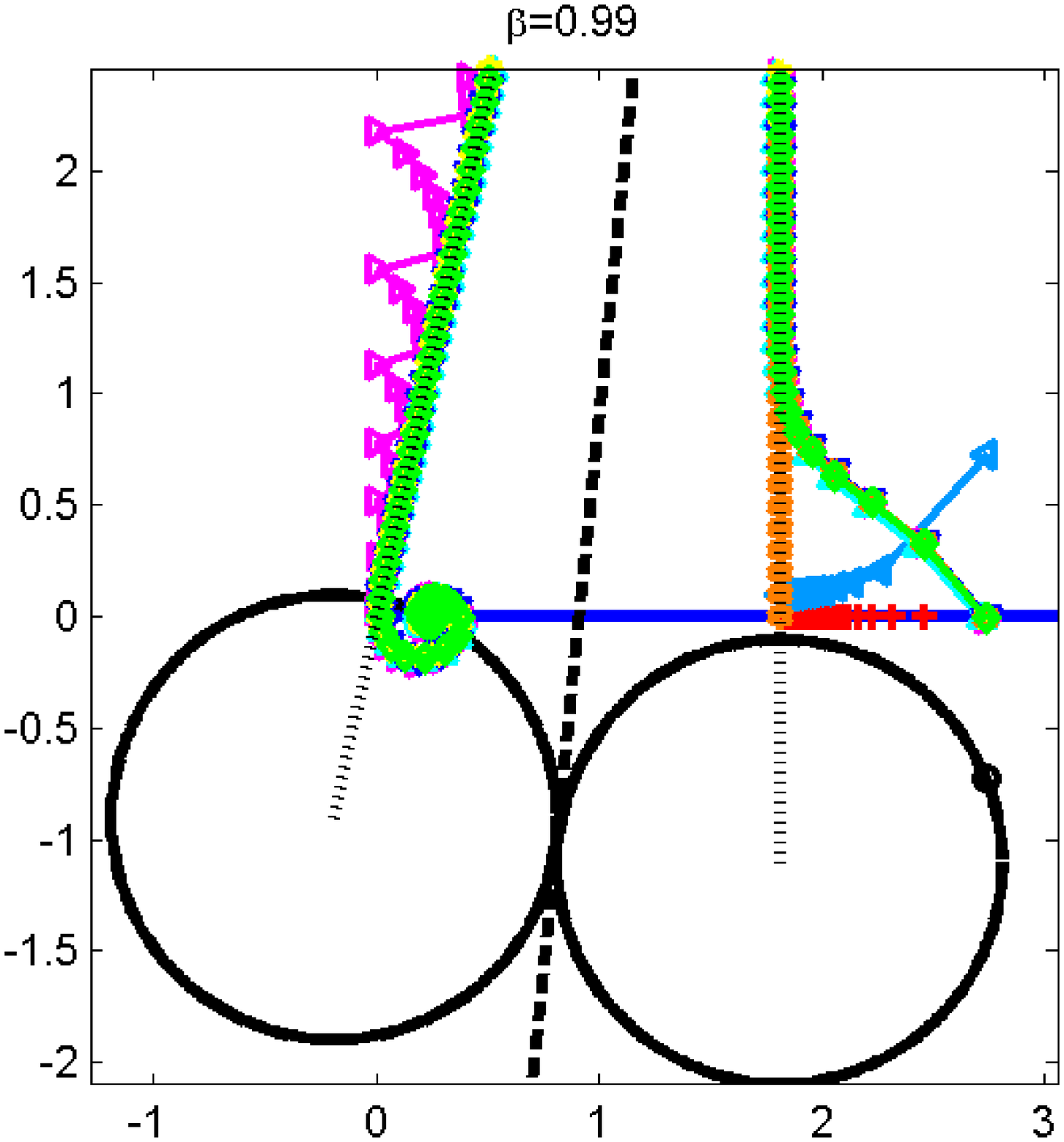}
}\\[-19pt]
\subfigure[]
    {\label{2circlesposb}
\includegraphics[width=0.35\textwidth]{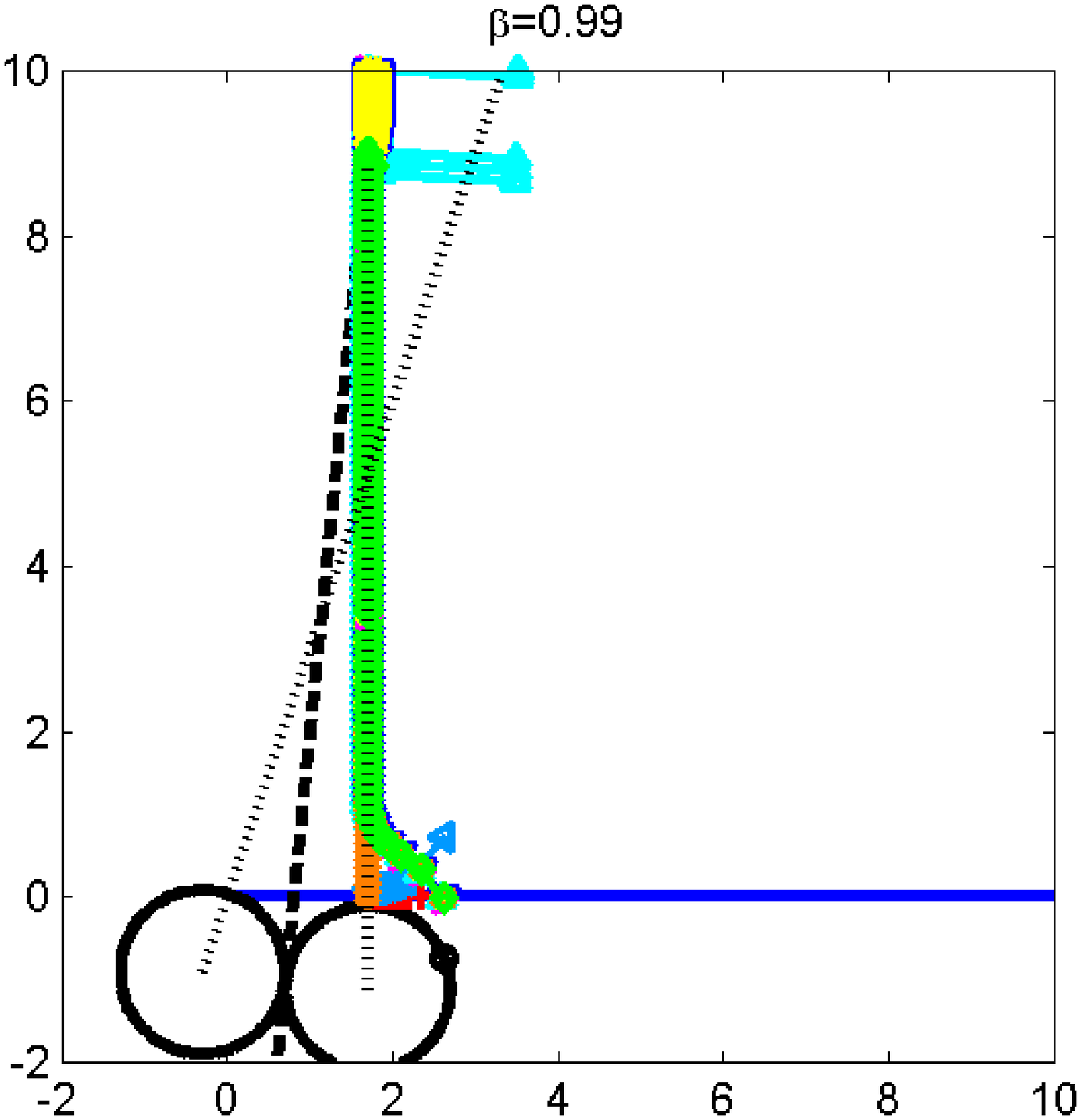}
	\usebox{\mylegend}
\includegraphics[width=0.35\textwidth]{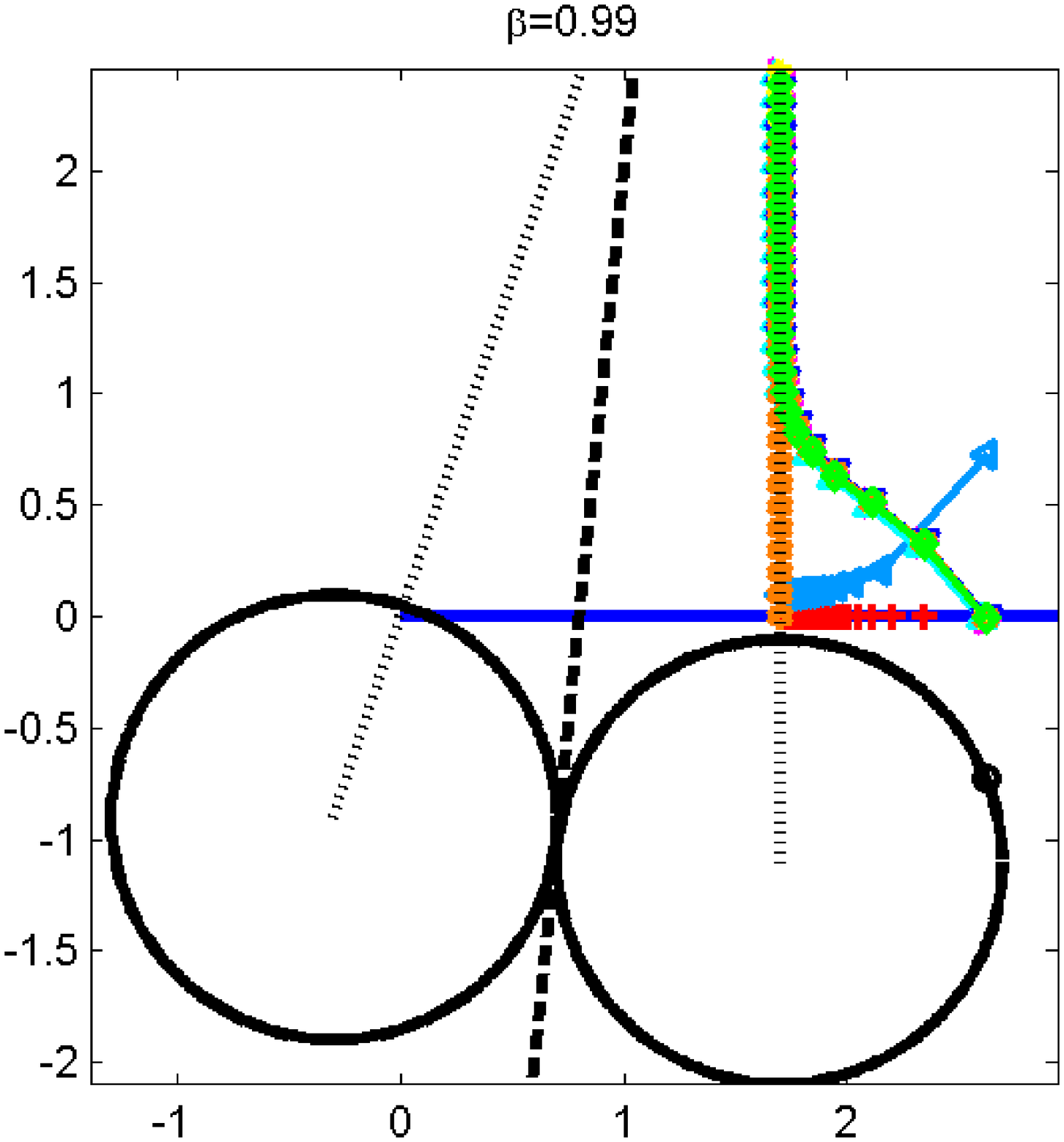}
}\\[-19pt]
\subfigure[]
	{\label{2circlesposc}
\includegraphics[width=0.35\textwidth]{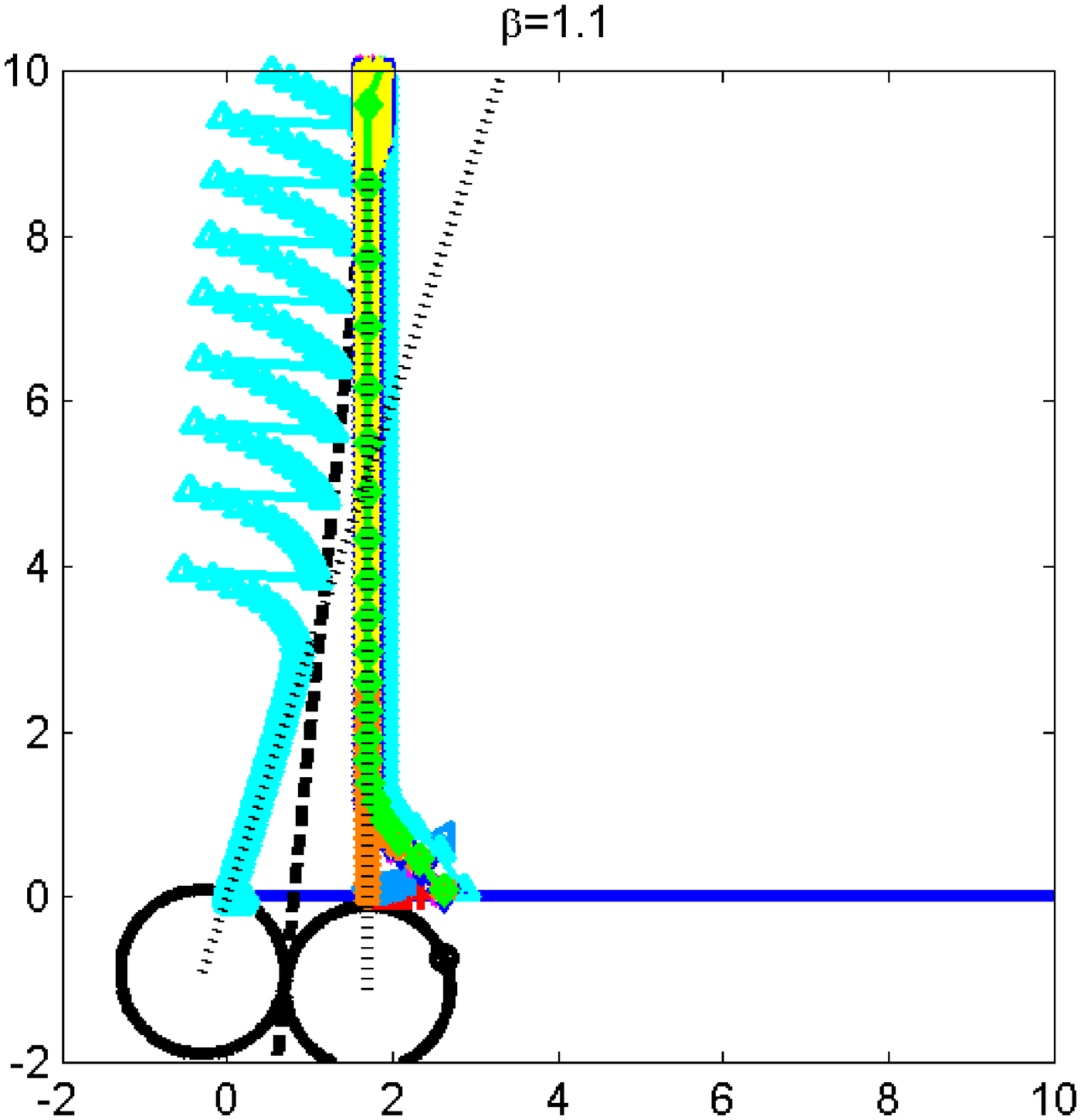}
	\usebox{\mylegend}
\includegraphics[width=0.35\textwidth]{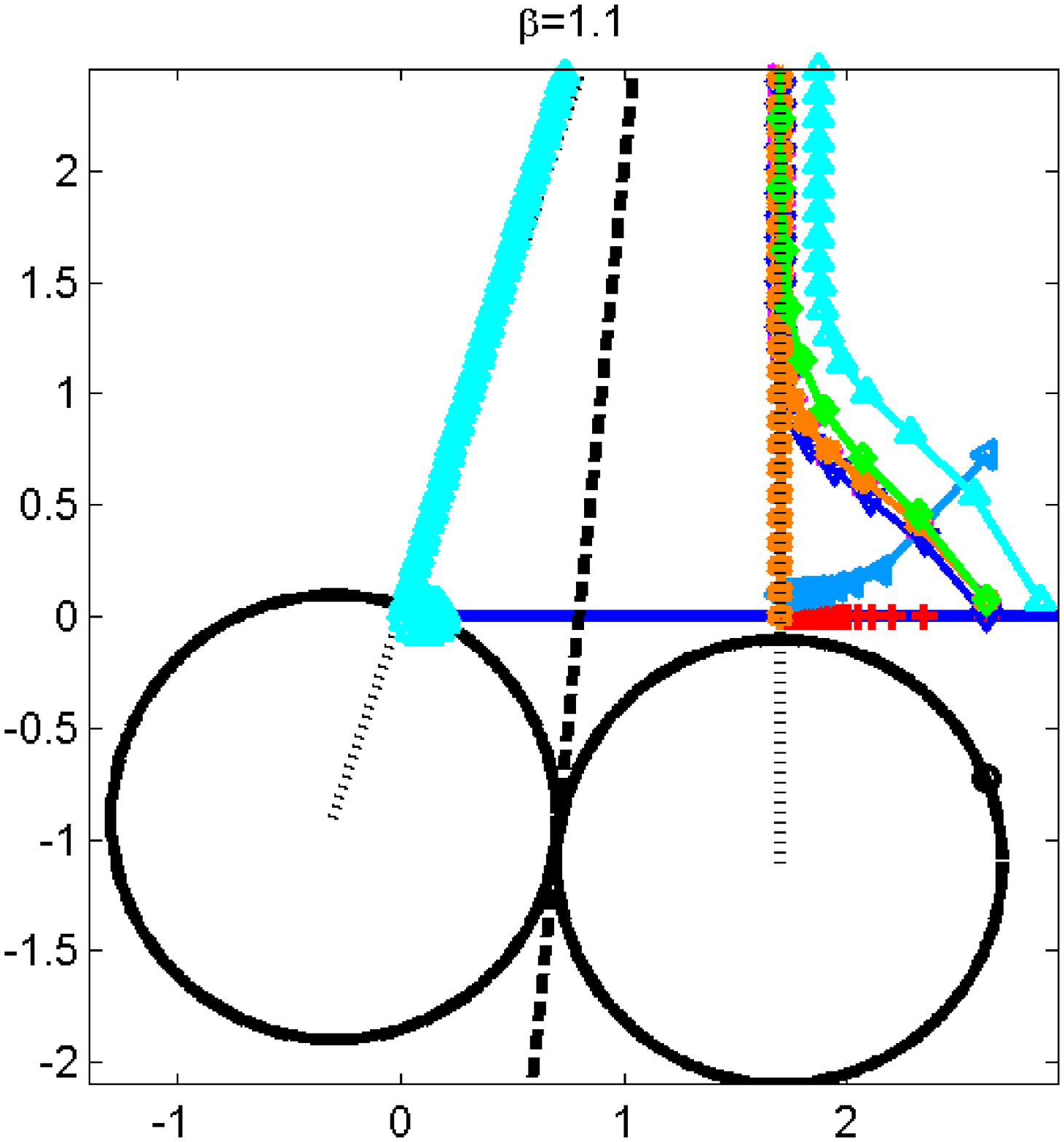}
}\\[-10pt]
\caption{(a) 
The starting point is again on the circle to the right, close to a 
local minimum. HIO and variant move away from the local minimum in 
the direction of the gap untill they reach the region where the circle 
to the left is closer. 
Instead of moving in a spiral like fashon, the iterations move close to 
the dotted line joining the center of the left circle to the origin, 
except for HIO that bounches on the x=0 axis.
(b) The solution is very close to 0, and the dotted line 
originating from the circle to the left and passing
 by the origin becomes more tilted. The various algorithms after moving 
in the vertical direction away from the local minimum, reach the dashed line
and start moving toward the tilted dotted line, falling back in the region closer to the first minimum. Tthese algorithms bounce between the 
regions closer to each circle without reaching the solution.
With $\beta>1$, i.e. inverting the order of the operators, Diff. Map 
converges, and RAAS diverges, while HIO HPR and ASR stagnate.}
\label{2circlespositivity}
\end{figure*}

The situation changes slightly when we consider the positivity constraint. 
The previous definitions of the algorithms still apply just replacing 
$\bm{P}_S$ with $\bm{P}_{S+}$:
\begin{equation}
\bm{P}_{S+}=\begin{cases}
 \rho(x)\ &\text{if $x \in S$ and $\rho(x)\ge 0$} \\
0 &\text{otherwise.}
\end{cases}
\end{equation}
 The only difference is HIO which becomes:
\begin{equation}
 \rho^{(n+1)}=
\begin{cases}
\bm{P_m} \rho^{(n)}(x)  & 
	\text {if  $x\in S$ and $\bm{P_m} \rho^{(n)}(x)\ge 0$} \\
(1-\beta \bm{P_m})\rho^{(n)}  & \text{otherwise.}
\end{cases}
\end{equation}
Fig. \ref{2circlesposa} shows that HIO bouches at the $x=0$ axis. As
the positivity constraint gets closer to the solution, none of the
algorithms converges to the solution (Fig. \ref{2circlesposb}), with 
the HIO-type algorithms bouncing between the regions closer 
to the two circles.
Only Difference Map for $\beta>1$ converges (Fig. \ref{2circlesposc}). 
Also HIO+ER would reach the solution for larger
intervals between ER iterations.

\section{Conclusions}
ER is a simple but powerfull local minimizer, HIO and variants are
very powerfull in escaping local minima, but in several situations
fail to converge. When positivity is introduced, the recursive 
version of HIO (HPR) converges more
`smoothly' to the solution without bounching on the $x=0$ axis.
Alternating between HIO and ER with the correct
intervals would have worked in all the examples shown above. 
 RAAS is a good (single parameter) way to change from `global' to local
minimizer, although it seems better to start from a high value of $\beta$
and decrease it afterwards.
Difference Map is succesfull in a few more of the examples shown above for
the proper choiche of $\beta$, however it involves 2 time consuming
modulus constraint operations. 
To find when stagnation occours one can monitor the distances
(using the proper error metrics) of the current solution before and
after aplying various projectors, 
or  monitor the autocorrelation between two succesive reconstructions. 
SPEDEN, a conjugate gradient based method, reaches a local minimum 
with quadradic convergence, and provides such kind of information.

The Solvent flipping algorithm does not show much success in the
examples shown above. Despite this it was used to improve
images \cite{abrahams:1996}, and in a modified form to solve
3D structures ab-initio \cite{chargeflip}. Perhaps the reason for the
latter success has more to do with the {\it threshold} constraint
used. A \textit{threshold projector} multiplies by 0 everything that is
below a given threshold, while a \textit{threshold reflector} multiplies 
it by -1. 
The application of this constraint has been proven succesfull in obtaining 
{\it ab-initio} solution in many
circumstances applied in a variety of ways. 
Apart from Charge Flipping, 
which uses a threshold reflector \cite{chargeflip}, 
in electron density modification procedure with SIR
\cite{giacovazzo:preprint},  in atomicity constraint 
\cite{Elser:acta}, a modified histogram constraint using the Difference Map
(H. He, private comm.),
and for complex valued objects, in the shrink-wrap algorithm using an
updated support constraint by thresholding the current object 
reconstruction at low resolution \cite{marchesini:prbr}. 

By compressing the image in small spots and by increasing the flat region, 
these algorithms based on the threshold constraint 
resemble the maximum entropy method \cite{maxent}, which 
tries to reduce the amout of information in an image and maximize 
the number of pixels of equal value (solvent).

What distinguishes shrink-wrap and SIR algorithms is that they somehow 
solve the low resolution first, and gradually introduce high resolution
information while slowly updating the low resolution information. 
Fienup has also shown that starting from a
low resolution image, slowly increasing resolution improves the algorithm
 \cite{fienup:1990}. Also SPEDEN starting from a low resolution 
target has been shown to slowly extend
 the correct phase information to higher resolution.
One possible explanation could be that the set based on
low resolution is `more smooth'. Perhaps also slowly increasing the gray 
levels in an image, or the number of possible phases of its Fourier transform
 could improve convergence by gradually introducing 
more degrees of freedom in the resulting image, although 
a set defined by `quantized' levels is also non-convex rendering it 
counterproductive. 

\begin{acknowledgments}
This work was performed under the auspices of the U.S. Department of 
Energy by the Lawrence Livermore National Laboratory under Contract 
No. W-7405-ENG-48 and the Director, Office of Energy Research. 
This work was funded by the  National Science Foundation through the Center
 for Biophotonics. 
The Center for Biophotonics, a National Science Foundation
 Science and Technology Center, is managed by the University of 
California, Davis, under Cooperative 
Agreement No. PHY0120999. D.~R.~Luke provided very usefull comments.
\end{acknowledgments}


\begin{thebibliography}{27}
\bibitem{Gerchberg:1972} R.~Gerchberg and W.~Saxton, Optik 
\textbf{35}, 237 (1972).
\bibitem{fienup:1978} J.~R.~Fienup, \ol \textbf{3}, 27-29 (1978).
\bibitem{fienup:1982} J.~R.~Fienup, \ao \textbf{21}, 
2758 (1982).
\bibitem{cederquist:1988} J.~N.~Cederquist, J.~R.~Fienup, 
J.~C.~Marron, R.~G.~Paxman, \ol \textbf{13}, 619. (1988).
\bibitem{stark:1984} A.~Levi and H.~Stark, \josaa \textbf{1},
932-943 (1984).
\bibitem{stark:1987} H.~Stark, \textit{Image Recovery: Theory and
 applications}. (Academic Press, New York, 1987).
\bibitem{elser:2003} V.~Elser, \josaa \textbf{20}, 40 (2003). 
\bibitem{luke:1}  H.~H.~Bauschke, P.~L.~Combettes, and
D.~R.~Luke. \josaa \textbf{19}, 1334-1345 (2002).
\bibitem{luke:2}  H.~H.~Bauschke, P. L. Combettes, and D. R. Luke,
\josaa  \textbf{20}, 1025-1034 (2003).
\bibitem{luke:3} D.~R.~Luke, (2003) [PIMS-03-13].
\bibitem{speden}  S.~P.~Hau-Riege, H.~Sz\"oke, H.~N.~Chapman et al. (2004) [arXiv:physics.optics/0403091]
\bibitem{luke:siam} D.~R.~Luke, J. V. Burke, R. G. Lyon, 
SIAM Review \textbf{44} 169-224 (2002).
\bibitem{luke:siam1} D.~R.~Luke, J. V. Burke, R. G. Lyon, SIAM 
J. Contr. Opt. \textbf{42}, 576-595 (2003).
\bibitem{bregman:1965} L.~M.~Br\`egman, Sov. Math. Dokl. \textbf{6}, 
688-692 (1965).
\bibitem{abrahams:1996}  J.~P.~Abrahams, A.~W.~G.~Leslie, Acta Cryst. 
\textbf{D52}, 30-42 (1996)
\bibitem{Elser:acta} V. Elser, Acta Cryst. \textbf{A59}, 201-209 (2003), 
[arXiv:cond-mat/0209690].
\bibitem{chargeflip} G. Oszl\'anyi and A. S\"ut\H o, Acta Cryst.  \textbf{A60}, 134-141 (2004) [arXiv:cond-mat/0308129]
\bibitem{giacovazzo:preprint} B. Carrozzini, G. L. Cascarano, L.De Caro, et al.
[arXiv:physics.optics/0404073]
\bibitem{marchesini:prbr} S.~Marchesini et al. \prb \textbf{68}, 140101(R) (2003) [arXiv:physics.optics/0306174]
\bibitem{maxent} G.~Bricogne, Acta Cryst. \textbf{A44}, 517-545 (1988).
\bibitem{fienup:1990} J.~R.~Fienup, A.~M.~Kowalczyk  
\josaa \textbf{ 7}, 450 (1990).
\end{thebibliography}
\end{document}